\documentclass[10pt,english,floatfix,nofootinbib,superscriptaddress,aps,prd,preprint,twocolumn,showkeys]{revtex4}
\usepackage[utf8]{inputenc}
\usepackage{float}
\usepackage{array}
\usepackage{lipsum}
\usepackage{bbm}
\usepackage{dsfont}
\usepackage{graphicx}
\usepackage{amsmath}
\usepackage{tikz}
\usepackage{multirow}
\usepackage{braket}
\usepackage{bbm}
\usetikzlibrary{quotes,angles}
\usetikzlibrary{arrows} 
\usetikzlibrary{decorations.markings}
\usepackage{graphicx}
\usepackage[english]{babel}
\usepackage{color}
\usepackage{subfig}
\usepackage{caption}
\usepackage{tensor}
\usepackage{booktabs}
\usepackage{esint}
\usepackage[dvips]{epsfig}
\usepackage[dvips]{graphicx}
\usepackage{float}
\usepackage{units}
\usepackage{textcomp}
\usepackage{mathrsfs}
\usepackage{amsmath}
\usepackage[makeroom]{cancel}
\usepackage{amssymb}
\usepackage{amsbsy}
\usepackage{amsfonts}
\usepackage{amssymb,mathrsfs,xcolor}
\usepackage{esint}
\usepackage{array}
\usepackage{graphicx}
\usepackage{hyperref}
\usepackage{enumerate}
\hypersetup{
    colorlinks,breaklinks,
    citecolor=[rgb]{0,0.0,1.0},
    urlcolor=[rgb]{0.0,0.0,1.0},
    linkcolor=[rgb]{0,0.5,0.9}}
\usepackage{slashed}
\usepackage{hyperref}

\begin{document}

\title{The accretion disk and neutrino propagation of Barrow-modified Black Hole}

\author{Yuxuan Shi}
\email{shiyx2280771974@gmail.com}
\affiliation{Department of Physics, East China University of Science and Technology, Shanghai 200237, China}

\author{Hongbo Cheng}
\email{hbcheng@ecust.edu.cn}
\affiliation{Department of Physics, East China University of Science and Technology, Shanghai 200237, China}
\affiliation{The Shanghai Key Laboratory of Astrophysics, Shanghai, 200234, China}

\begin{abstract}

This paper attempts to clarify the deep consequences of Barrow fractal black hole spacetime configurations caused by quantum gravity on neutrino pair annihilation and accretion disk dynamics. We systematically derive the analytical expression for the innermost stable circular orbit (ISCO) radius ($r_{\text{ISCO}}\propto M^{2/(2+\Delta)}$) by building a Barrow-modified static spherically symmetric metric ($r\rightarrow r^{1+\Delta/2}$), and we find that increasing $\Delta$ significantly shifts the ISCO inward. We numerically solve the radiation flux, effective temperature, and differential luminosity distribution under the modified metric based on the Novikov-Thorne relativistic thin accretion disk model. For $\Delta=1$, the results show that the temperature increases by $62.5\%$, the peak disk radiation flux increases by $22.5\%$, and the spectral radiance increases by around $50\%$. Fractal horizons enhance neutrino trajectory bending effects, according to further study of neutrino pair annihilation ($\nu\bar{\nu}\rightarrow e^+e^-$) energy deposition processes using local Lorentz transformations and null geodesic equations. The energy deposition rate for $\Delta=1$ is $8-28$ times higher than classical estimates when the black hole radius is $R/M\sim3-4$. This work provides important theoretical insights into the influence of quantum spacetime geometry on high-energy astrophysical phenomena in extreme gravitational fields by establishing, for the first time, quantitative relationships between the Barrow parameter $\Delta$ and neutrino pair annihilation energy and accretion disk radiative efficiency.

\end{abstract}

\maketitle

%\tableofcontents

\section{Introduction}

General relativity, which was established by Einstein, has produced major advances in cosmology by explaining and forecasting several occurrences in the study of the cosmos \cite{chandrasekhar1998mathematical,weinberg2013gravitation}. The thermodynamic characteristics \cite{hawking1983thermodynamics,witten1998anti,chamblin1999holography,bekenstein1973black,ryu2006aspects} and high-energy physical processes \cite{abbott2020properties,gewirth1988electronic, arons2003magnetars} in strong gravitational fields of black holes, one of the most difficult astrophysical objects predicted by general relativity, continue to be major subjects of astrophysics study. The fundamental framework for black hole thermodynamics was created by the conventional Bekenstein-Hawking entropy \cite{hawking1983thermodynamics, bekenstein1973black}, which was based on the event horizon area $A_+$. However, the geometric structure of black hole horizons may be profoundly changed by quantum gravitational processes. The fractal horizon model proposed by Barrow \cite{barrow2020area} offers a fresh perspective: the surface of a black hole may exhibit complex layered patterns at minuscule scales by adding a fractal dimension parameter $\Delta\,(0\leq\Delta\leq1)$. This remarkably expands the effective horizon area and modifies the entropy to $S_B\propto A_+^{1+\Delta/2}$. Through quantum adjustments to the metric (horizon radius $r\to r^{1+\Delta/2}$ \cite{petridis2023barrow}), this correction not only affects the thermodynamic development of black holes themselves but also changes the dynamics of exterior spacetime.

The thermodynamic characteristics of Barrow-modified black holes and their cosmological uses have been the main subjects of previous research \cite{barrow2020area,petridis2023barrow,capozziello2025barrow,wang2022barrow,abreu2020barrow,abdalla2022cosmology,nojiri2022early,saridakis2020barrow,drepanou2022kaniadakis,lu2025generalized,di2022sign,xia2024upper}. In the gravitational field equation, Barrow entropy is equivalent to the scaled cosmological constant \cite{lu2025generalized,di2022barrow}. Fractal horizons considerably extend the black holes' evaporation period and cause a positive qualitative shift in heat capacity, according to thermodynamic studies \cite{barrow2020area}. This suggests that there may be stable remnant black holes. The holographic dark energy model was modified to include Barrow entropy at the cosmic stage, which effectively accounted for the late universe's fast expansion \cite{saridakis2020barrow, anagnostopoulos2020observational, moradpour2020generalized, adhikary2021barrow}. These studies, however, are mainly restricted to large-scale cosmological phenomena or black holes proper, and systematic research on the effects of the Barrow-corrected spacetime geometry on high-energy astrophysical processes surrounding black holes, such as accretion disk radiation and neutrino pair annihilation process, is still lacking.

Despite its significant cosmological applications, fractal spacetime has not yet been thoroughly investigated for its impact on accretion disk radiation in high gravity fields. As the main sources of mass accretion onto black holes, accretion disks \cite{zuluaga2021accretion,page1974disk,di2002neutrino, janiuk2013accretion,kawanaka2007neutrino} have radiative characteristics that are intimately related to effective potential distributions and the innermost stable circular orbit. The temperature profile, radiative efficiency, and luminosity properties of the disk may be affected by changes in spacetime curvature, which might reinterpret the stability criteria of accretion fluxes under the Barrow-modified metric. Yet, under several gravitational conditions, neutrino-antineutrino pair annihilation is an essential source of energy for high-energy events like active galactic nucleus jets and gamma-ray bursts \cite{salmonson1999general,asano2000neutrino,asano2001relativistic,miller2003off,birkl2007neutrino,lambiase2020effects,prasanna2002energy,lambiase2021neutrino,shi2022neutrino,shi2023gamma,shi2024shadow}. According to the Ref.\cite{di2022sign}, fractal spacetime may have an impact on extreme astrophysical events like gamma-ray burst energy deposition. It does this by explaining the sign switching of the cosmological constant using the Barrow index. Given that spacetime geometry bends neutrino paths, metric corrections brought about by fractal horizons may greatly enhance energy deposition rates, a mechanism that has not yet been well investigated in Barrow-modified black hole models. 

This work examines the physics of neutrino pair annihilation processes in Barrow-modified black holes and accretion disks, methodically examining how the fractal parameter $\Delta$ affects the following important questions: 
\begin{enumerate}[(i)]
\item ISCO radii's evolution under modified metrics and their disk structure constraints;
\item accretion flow radiative signatures, temperature profiles, and effective potentials that depend on $\Delta$;
\item Barrow fractal sapcetime corrections to neutrino trajectory bending and energy deposition rates.
\end{enumerate}
Through the construction of null geodesic models and self-consistent hydrodynamic equations, our goal is to clarify how fractal horizons modify macroscopic behaviours through microscopic spacetime geometry, offering fresh perspectives on the function of quantum gravity in high-energy processes of compact objects. In the paper, we choose the units $c=\hbar=G=\ell_p=1$.

\section{Barrow structure and dynamical equations}

One of the first fractals, the Koch snowflake, was developed by Helge von Koch \cite{von2019continuous} in 1904 and allows for the construction of a two-dimensional structure with an infinite perimeter and finite area through a series of iterative steps. The Menger Sponge \cite{menger2002allgemeine} and the Sierpinski Gasket \cite{sierpinski1915curve}, which have infinite surface area and finite volume, are three-dimensional representations of this structure. Inspired by von Koch's theories, John Barrow has suggested that the surface of static, spherically symmetric Schwarzschild black holes with finite volume might exhibit a variety of intricate fractal formations that trend to infinity. Barrow postulated that the event horizon of the black hole around the contact sphere contains N tiny, stacked spheres with radius $\lambda$ times smaller than the initial sphere. After countless steps, the volume and area of the black hole, given a hierarchical structure with radius $r_{n+1} = \lambda r_{n}$, are \cite{barrow2020area}
\begin{align}
V_{\infty}&=\dfrac{4\pi}{3}r_+^3\sum_{n=0}^{\infty}\left(N\lambda^3\right)^n=\dfrac{4\pi r_+^3}{3\left(1-N\lambda^3\right)},\\
A_{\infty}&=4\pi r_+^2\sum_{n=0}^{\infty}\left(N\lambda^2\right)^n=\dfrac{4\pi r_+^2}{1-N\lambda^2},
\end{align}
where $r_+=2M$ is the Schwarzschild radius of the black hole and $M$ is the black hole mass. Therefore, the surface area will be unlimited but the volume will be finite if
\begin{align}
\lambda^{-2}<N<\lambda^{-3}.
\end{align}
For Schwarzschild radius $r_+$, the entropy is defined as \cite{hawking1983thermodynamics}
\begin{align}
\label{entropySch}
S=\pi\left(\dfrac{A_+}{A_pl}\right),
\end{align}
where the Planck area is $A_{pl}=4\pi\ell_p^2\sim4\pi$. According to the Barrow statistics, the area is
\begin{align}
A_{\infty}=\dfrac{A_+}{1-N\lambda^2}.
\end{align}
So that, $\Delta$ is used to introduce the entropy \cite{barrow2020area},
\begin{align}
\label{entropyBarrow}
S_B=\pi\left(\dfrac{A_+}{A_{pl}}\right)^{1+\frac{\Delta}{2}}.
\end{align}
The following modifications on $r$ may be produced by comparing Eq.(\ref{entropyBarrow}) and Eq.(\ref{entropySch}) \cite{petridis2023barrow},
\begin{align}
r\to r^{1+\frac{\Delta}{2}}.
\end{align}
It is possible to express the Barrow-modified metric as \cite{petridis2023barrow}
\begin{align}
\label{metric}
\mathrm{d}s^2&=g_{\mu\nu}\mathrm{d}x^{\mu}\mathrm{d}x^{\nu}\notag\\
&=-\left(1-\dfrac{2M}{r^{1+\frac{\Delta}{2}}}\right)\mathrm{d}t^2\notag\\
&\quad+\left(1-\dfrac{2M}{r^{1+\frac{\Delta}{2}}}\right)^{-1}\left(1+\dfrac{\Delta}{2}\right)r^{\Delta}\mathrm{d}r^2\notag\\
&\quad+r^{2+\Delta}\left(\mathrm{d}\theta^2+\sin^2\theta\mathrm{d}\varphi^2\right),
\end{align}
with $\Delta=0$, this leads to the classical case corresponding to the simplest horizon structure. At the most complex, $\Delta = 1$, it acts as though it had one more geometric dimension even though it is an area from an information standpoint. Subsequently, limited to the equatorial plane $(\theta = \pi/2)$, the Lagrangian is expressed as follows \cite{salmonson1999general},
\begin{align}
\label{lagrangian}
\mathcal{L}=\dfrac{1}{2}\left(g_{tt}\dot{t}^2+g_{rr}\dot{r}^2+g_{\varphi\varphi}\dot{\varphi}^2\right),
\end{align}
where the dots indicate the derivative with respect to an affine parameter $\tau$ that may be interpreted as the correct time for massive particles that follow timelike geodesics. By introducing the Lagrange equation, 
\begin{align}
\dfrac{\mathrm{d}}{\mathrm{d}\tau}\dfrac{\partial\mathcal{L}}{\partial\dot{x}^{\mu}}-\dfrac{\partial\mathcal{L}}{\partial x^{\mu}}=0,
\end{align}
the generalized momenta are
\begin{align}
\label{pt}
p_t&\equiv\dfrac{\partial\mathcal{L}}{\partial\dot{t}}=-\left(1-\dfrac{2M}{r^{1+\frac{\Delta}{2}}}\right)\dot{t}=-E,
\end{align}
\begin{align}
\label{pphi}
p_{\varphi}&\equiv\dfrac{\partial\mathcal{L}}{\partial\dot{\varphi}}=r^{2+\Delta}\dot{\varphi}=L.
\end{align}
For the particle describing the trajectory, the constants $E$ and $L$ stand for energy and angular momentum per unit rest mass, respectively. The particle is characterized by the equation of radial motion with momenta (\ref{pt}) and (\ref{pphi}) as follows, moving along the planar geodesics in the equatorial plane guided by the metric (\ref{metric}) \cite{heydari2023thin},
\begin{align}
g_{rr}\dot{r}^2=V_{\text{eff}}(r),
\end{align}
where the effective potential is \cite{heydari2023thin}
\begin{align}
\label{Veff}
V_{\text{eff}}(r)=\dfrac{E^2g_{\varphi\varphi}+L^2g_{tt}}{-g_{tt}g_{\varphi\varphi}}-1,
\end{align}

\begin{figure}
\centering
\includegraphics[scale=0.55]{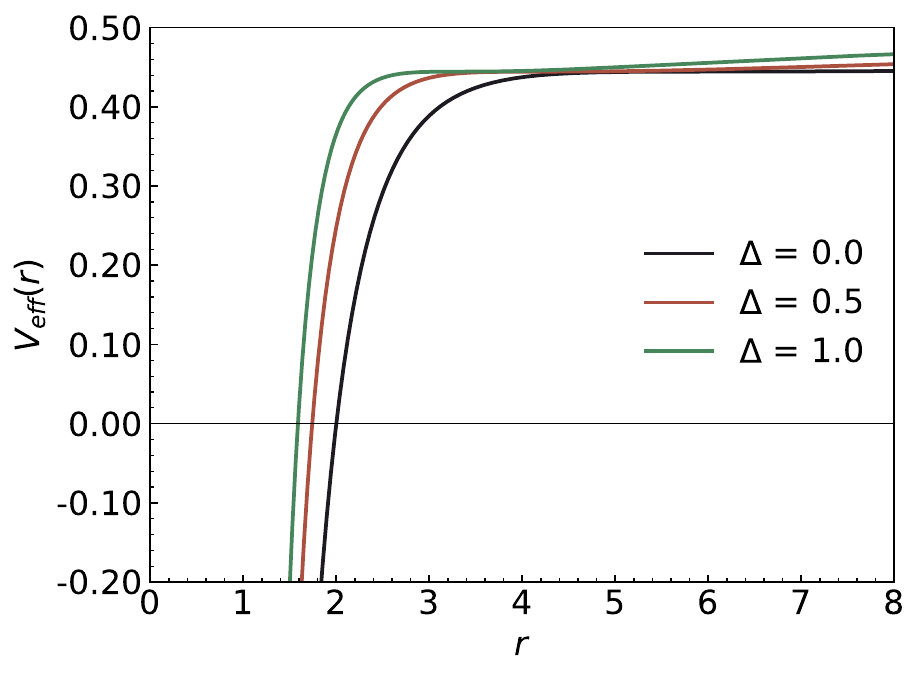}
\caption{For $\Delta = 0, 0.5$, and $1$, the effective potential is shown as a function of radial coordinate, accordingly.}
\label{fig_Veff}
\end{figure}

In Fig.\ref{fig_Veff}, the effective potential  is displayed. A set of potential curves are comparable. It should be noted that there are zeros in the effective potential, and that as $\Delta$ increases, the zero values diminish. According to the geodesic equation,
\begin{align}
\label{geo}
\dfrac{\mathrm{d}}{\mathrm{d}\tau}\left(g_{\mu\nu}\dot{x}^{\nu}\right)=\dfrac{1}{2}\left(\partial_{\mu}g_{\mu\nu}\right)\dot{x}^{\nu}\dot{x}^{\rho},
\end{align}
at the circle orbit in the equatorial plane, $\dot{r}=\dot{\theta}=\ddot{r}=0$, Eq.(\ref{geo}) can be reduced as
\begin{align}
\label{geo2}
\left(\partial_rg_{tt}\right)\dot{t}^2+\left(\partial_rg_{\varphi\varphi}\right)\dot{\varphi}^2=0.
\end{align}
The angular velocity can be calculated from Eq.(\ref{geo2}) as \cite{bambi2017black}
\begin{align}
\label{omega}
\Omega&=\dfrac{\dot{\varphi}}{\dot{t}}\notag\\
&=\sqrt{-\dfrac{\partial_rg_{tt}}{\partial_rg_{\varphi\varphi}}}.
\end{align}
Therefore, from Eq.(\ref{omega}), we obtain an expression for the particular angular momentum \cite{harko2009testing}
\begin{align}
\label{momentum}
L&=\dfrac{\Omega g_{\varphi\varphi}}{\sqrt{-g_{tt}-\Omega^2g_{\varphi\varphi}}}\notag\\
&=\dfrac{\sqrt{M}r^{\frac{2+\Delta}{4}}}{\sqrt{1-3Mr^{-1-\frac{\Delta}{2}}}},
\end{align}
and energy \cite{harko2009testing}
\begin{align}
\label{energy}
E&=-\dfrac{g_{tt}}{\sqrt{-g_{tt}-\Omega^2g_{\varphi\varphi}}}\notag\\
&=\dfrac{1-2Mr^{-1-\frac{\Delta}{2}}}{\sqrt{1-3Mr^{-1-\frac{\Delta}{2}}}}.
\end{align}
The radial function $V_{\text{eff}}(r)$ is zero, just like its first derivative with regard to $r$,
\begin{align}
\partial_r V_{\text{eff}}(r)=\dfrac{E^2}{g_{tt}^2}\partial_r g_{tt}+\dfrac{L^2}{g_{\varphi\varphi}}\partial_r g_{\varphi\varphi}=0.
\end{align}
Moreover, the second-order derivative of the effective potential provides the stability condition, meaning that the orbiting particle's radius must meet $\partial_r^2 V_{\text{eff}}\leq0$ in order for the orbit to be stable. The innermost stable circular orbit (ISCO) is the equivalent radius when the second-order derivatives of the effective potential equal zero. The second-order derivative of $V_{\text{eff}}$ is written as
\begin{align}
\label{dVdr2}
\partial_r^2 V_{\text{eff}}&=\dfrac{E^2}{g_{tt}^3}\left[-2\left(\partial_r g_{tt}\right)^2+g_{tt}\partial_r^2 g_{tt}\right]\notag\\
&\quad+\dfrac{L^2}{g_{\varphi\varphi}^3}\left[-2\left(\partial_r g_{\varphi\varphi}\right)^2+g_{\varphi\varphi}\partial_r^2 g_{\varphi\varphi}\right].
\end{align}

\begin{figure}
\centering
\includegraphics[scale=0.55]{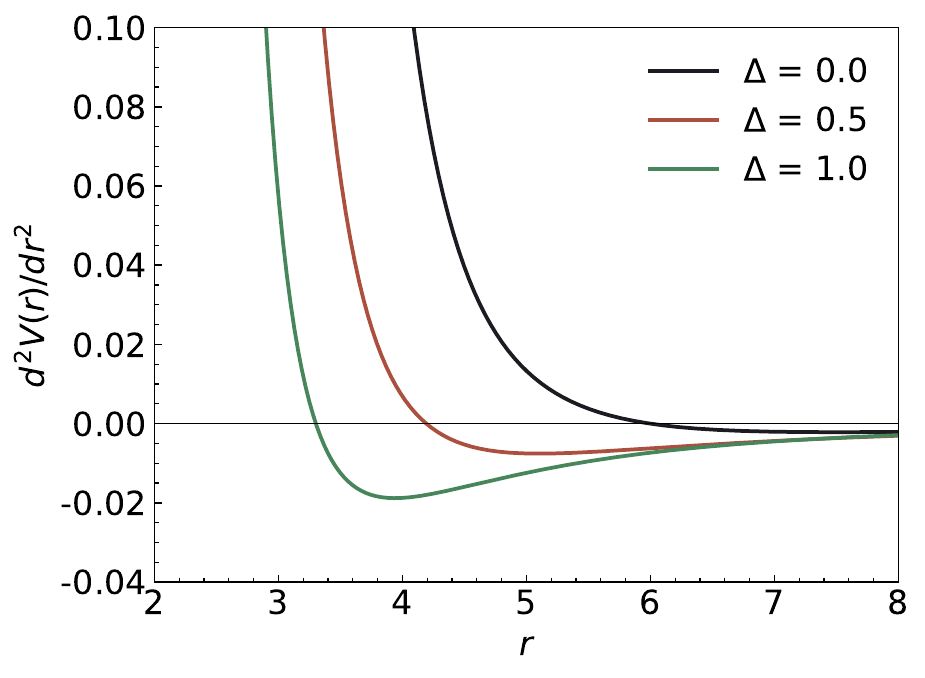}
\caption{Effective potential derivatives for $\Delta = 0, 0.5$, and $1$ are shown by curves, respectively.}
\label{fig_dVrdr2}
\end{figure}

We calculate the effective potential twice, and Fig.\ref{fig_dVrdr2} shows the derivative. With increasing $\Delta$, the zero values decrease. As the radial coordinate gets closer to infinity, the function tends to be around zero, according to Fig.\ref{fig_dVrdr2}. To determine the ISCO of Barrow BH, we set Eq.(\ref{dVdr2}) to zero, with applying Eqs.(\ref{metric}), (\ref{momentum}) and (\ref{energy}), to get the  following equation,
\begin{align}
M\left(6M-r^{1+\frac{\Delta}{2}}\right)\left(2+\Delta\right)^2=0,
\end{align}
whose real solution is
\begin{align}
\label{ISCO}
r_{\text{ISCO}}=36^{\frac{1}{2+\Delta}}M^{\frac{2}{2+\Delta}}.
\end{align}

\begin{figure}
\centering
\includegraphics[scale=0.55]{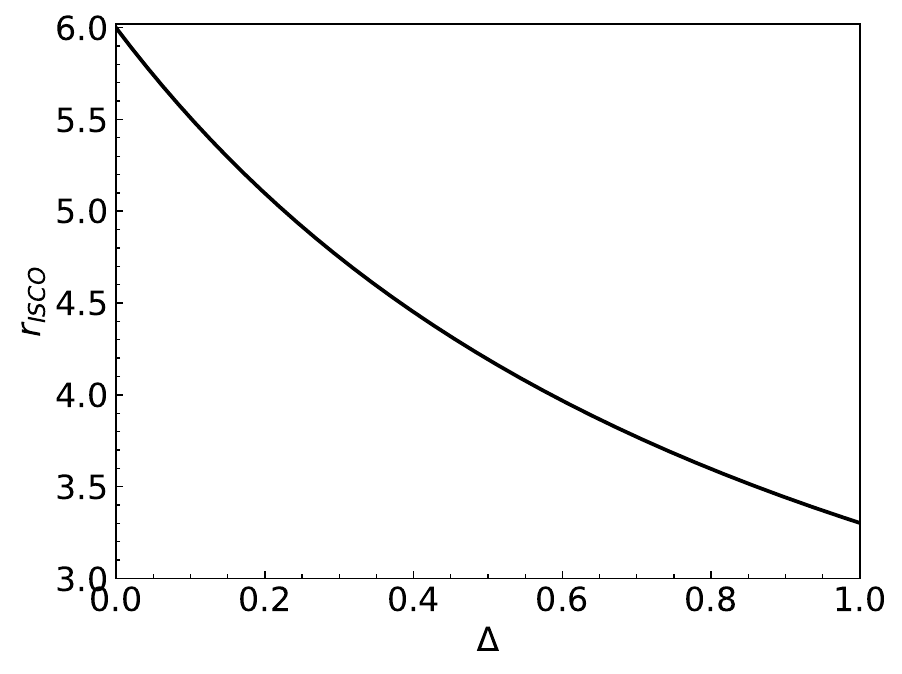}
\caption{The ISCO radius curve in relation to $\Delta$.}
\label{fig_ISCO}
\end{figure}

ISCO advances inward as $\Delta$ grows, as seen in Fig.\ref{fig_ISCO}, indicating that the critical point of orbital instability is heightened by the quantum geometric effect. The characteristic variable $\Delta$ is one of the factors that is related to the radius of ISCO by Eq.(\ref{ISCO}). In Fig.\ref{fig_ISCO}, the relationship is displayed. The lower the ISCO radius under the larger $\Delta$, the smaller the so-called ISCO radius, which is a decreasing function of the departure from the Barrow-modified black hole.

\section{Relativistic thin accretion disk}

The simplest non-relativistic model of an accretion disk around a compact core object assumes that turbulent viscosity moves angular momentum outward via the accretion disk while matter spirals inward, losing angular momentum in the process. The gas releases heat energy as it heats up and loses gravitational energy as it goes inside. A quasi-steady disk is assumed to be situated on the equatorial plane of a stationary, axisymmetric background spacetime geometry in the Novikov-Thorne model. The matter on the disk is assumed to travel in an orbit that approximates a geodesic circle. The disk is thin; that is, its greatest half-thickness, $H$, is $H/R\ll1$, where $R$ is the disk's characteristic radius \cite{page1974disk,novikov1973black, shakura1973black,thorne1974disk}. The averages of the azimuthal angle $\varphi = 2\pi$, the height $H$, and the time scale $\delta t$ of the inward flow of gas describe the thermal properties of the disk. The heat produced by stress and dynamic friction is efficiently emitted from the disk surface primarily in the form of radiation. Based on these presumptions The conservation law may be used to determine the disk's time-averaged radial structure under these presumptions. The conservation principles for energy, angular momentum, and the time-averaged radial structure of the disk may all be derived from the equation of conservation of rest mass. The consistency of the mass accretion rate is inferred from the integration of the equation of mass conservation \cite{bambi2017black},
\begin{align}
\dot{M}=-2\pi r\Sigma(r)u^r=\text{constant},
\end{align}
where $\Sigma(r)$ is the surface density and $u^r$ is the radial velocity.  The differential of the luminosity $\mathcal{L}_{\infty}$at infinity can be expressed as \cite{page1974disk, joshi2013distinguishing},
\begin{align}
\label{dLinf}
\dfrac{\mathrm{d}\mathcal{L}_{\infty}}{\mathrm{d}\ln r}=4\pi r\sqrt{-G}\mathcal{F}(r),
\end{align}
where \cite{bambi2012code}
\begin{align}
G=g_{tt}g_{tt}g_{\varphi\varphi}=\sqrt{r^{2+2\Delta}\left(1+\dfrac{\Delta}{2}\right)}.
\end{align}
In the local frame of the accreting fluid, the flow of radiant energy $\mathcal{F}$ released from the disk's top face is represented by the particular angular momentum $L$(\ref{momentum}), specific energy $E$(\ref{energy}), and angular velocity $\Omega$(\ref{omega}) \cite{page1974disk, thorne1974disk},
\begin{align}
\label{FMR}
\mathcal{F}(r)&=-\dfrac{\dot{M}}{4\pi\sqrt{-G}}\dfrac{\partial_r\Omega}{(E-\Omega L)^2}\notag\\
&\quad\times\int_{r_{\text{ISCO}}}^{r}(E-\Omega L)(\partial_r L)\mathrm{d}r.
\end{align}
The numerical integration of Eq.(\ref{FMR}) is eased by integrating by parts,
\begin{align}
&\quad\int_{r_{\text{ISCO}}}^{r}(E-\Omega L)(\partial_r L)\mathrm{d}r\notag\\
&=EL-(EL)_{\text{ISCO}}-2\int_{r_{\text{ISCO}}}^{r}L(\partial_r E)\mathrm{d}r.
\end{align}
Given that the disk is thought to be in thermodynamic equilibrium, the radiation released may be regarded as a black body radiation, with the temperature determined by,
\begin{align}
\label{Tr}
T(r)=\left[\dfrac{\mathcal{F}(r)}{\sigma}\right]^{\frac{1}{4}},
\end{align}
where $\sigma$ is the Stefan-Boltzmann constant.

The efficiency of converting accreted mass to radiation $\epsilon$ may be determined by measuring the energy loss of a test particle traveling from infinity to the disk's inner border, provided that all emitted photons escape to infinity. Taking into account that for $r\to\infty$ and $E_{\infty}\approx1$, we obtain
\begin{align}
\label{efficiency}
\epsilon=\dfrac{E_{\infty}-E_{\text{ISCO}}}{E_{\infty}}\approx1-\dfrac{2\sqrt{2}}{3}.
\end{align}
Taking into account photon trapping by the black hole, Eq.(\ref{efficiency}) displays the value of $\epsilon$, the efficiency with which the accreted mass of a Barrow-modified black hole is transformed into radiation. It is obvious that the value of $\epsilon$ is now a constant and has no relationship to the fractal parameter $\Delta$.

\begin{figure}
\centering
\includegraphics[scale=0.55]{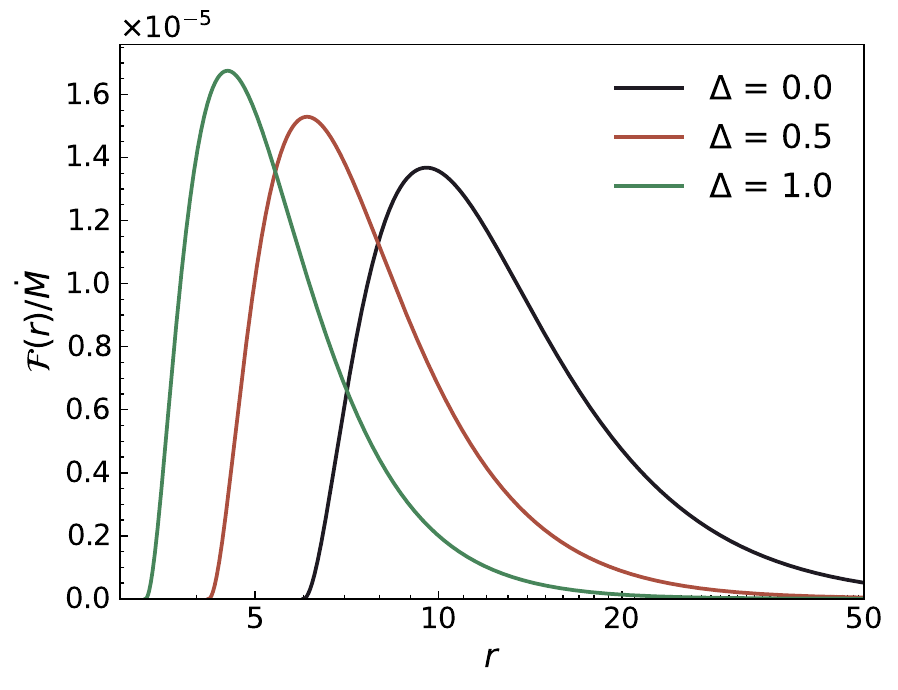}
\caption{Energy flux per unit accretion rate from a relativistic thin accretion disk around a Barrow-modified black hole for $\Delta=0, 0.5$ and $1$.}
\label{fig_FrdotM}
\end{figure}

\begin{figure}
\centering
\includegraphics[scale=0.55]{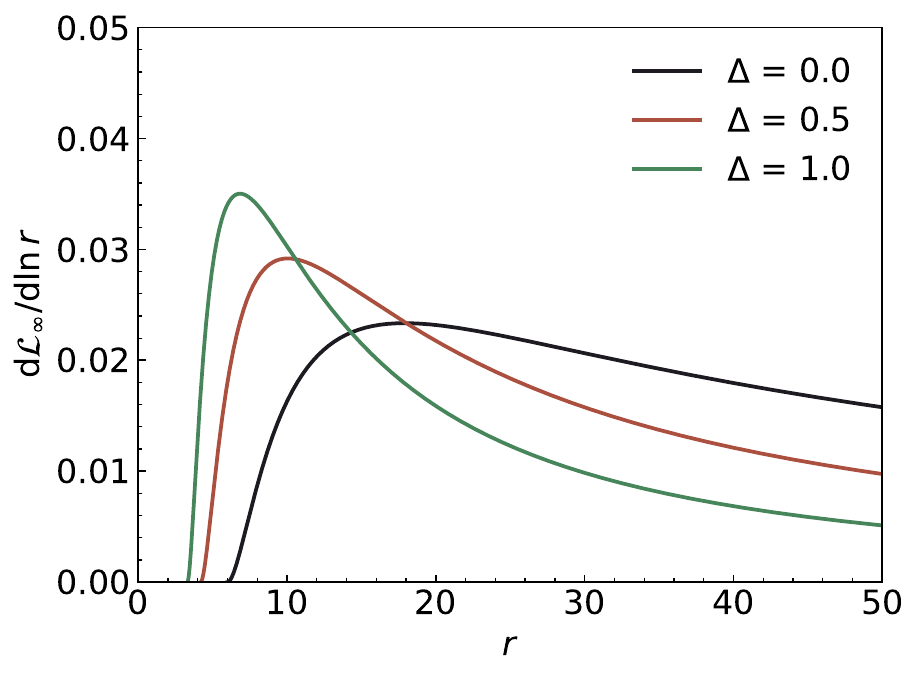}
\caption{The differential luminosity from a relativistic thin accretion disk around a Barrow-modified black hole for $\Delta=0, 0.5$ and $1$.}
\label{fig_dLinf}
\end{figure}

\begin{figure}
\centering
\includegraphics[scale=0.55]{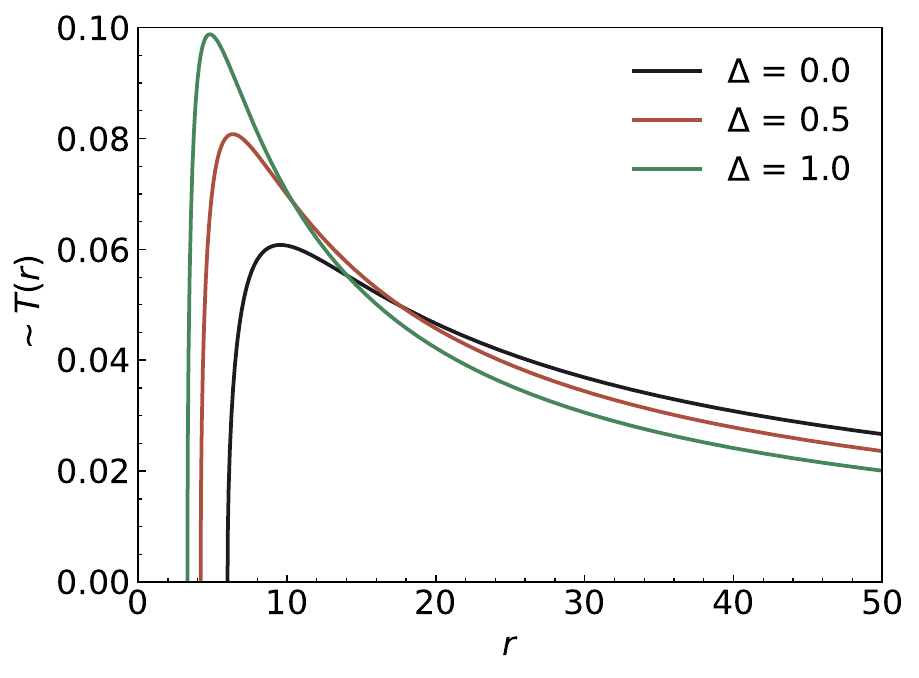}
\caption{Radial profiles of the temperature per unit accretion rate of a relativistic thin accretion disk around a Barrow-modified black hole for $\Delta=0, 0.5$ and $1$.}
\label{fig_Tr}
\end{figure}

The accretion disk surrounding the Barrow-modified black hole is presently being investigated. For the sake of comparison, we compute the mass accretion rate numerically in the following with $M = 1$. For the case of various fractal parameters $\Delta$, we present the curves for Eqs.(\ref{dLinf}), (\ref{FMR}), and (\ref{Tr}) in Figs.\ref{fig_FrdotM}, \ref{fig_dLinf} and \ref{fig_Tr}.

Fig.\ref{fig_FrdotM} shows that the disk of the Barrow-modified black hole can radiate more energy than in the classical scenario when Delta grows. The peak of the radiative flux's radial profile rises and shifts significantly inward as the disk's inner edge approaches lower values of $r$. Maximum value increases by $22.47\%$ for the most complicated fractal, $\Delta = 1$, and $11.8\%$ for $\Delta = 0.5$. The accretion disk's temperature and brightness similarly exhibit these effects, as seen in Figs.\ref{fig_dLinf} and \ref{fig_Tr}. When $\Delta=1$, the temperature rises by $62.47\%$ and the differential luminosity peak is about $\sim50\%$ greater than in the classical instances. Phase transitions can be triggered by the logarithmic adjustment of Barrow entropy for EMS black holes \cite{biswas2025einstein}. This, in conjunction with the accretion disk's temperature increasing by $62.5\%$ in Fig.\ref{fig_Tr}, supports the redistribution impact of fractal horizons on energy fluxes. It is noteworthy that the findings in Figs.\ref{fig_FrdotM}, \ref{fig_dLinf} and \ref{fig_Tr} demonstrate that the effects of quantum fractals extend beyond the black hole's field of view and even alter the thermal characteristics of the accretion disk surrounding the Schwarzschild black hole.

\section{The neutrino energy deposition}

In this section, we discuss $\nu\bar{\nu}\to e^+e^-$ energy deposition \cite{popham1999hyperaccreting,birkl2007neutrino,khodadi2023spontaneous,zalamea2011neutrino} around Barrow-modified black holes. Since a neutrino's mass is so tiny, it can be regarded as a massless particle, and the null geodesic equation can be used to describe its motion \cite{li2021shadows},
\begin{align}
\label{null_geo}
\left(\dfrac{\mathrm{d}r}{\mathrm{d}\varphi}\right)^2=\dfrac{g_{\varphi\varphi}^2}{g_{tt}}\left(\dfrac{1}{-g_{tt}b^2}-\dfrac{1}{g_{\varphi\varphi}}\right),
\end{align}
where $b=|L|/E$ is the impact parameter of the massive particle. It is possible to define the angle $\theta_r$ between the trajectory and the tangential velocity using the local longitudinal and radial velocities \cite{lambiase2020effects,prasanna2002energy},
\begin{align}
\label{theta_r}
\tan\theta_r=\sqrt{\dfrac{g_{rr}}{g_{\varphi\varphi}}}\dfrac{\mathrm{d}r}{\mathrm{d}\varphi}.
\end{align}
The relationship between $b$ and $\theta$ can be found by solving Eq.(\ref{null_geo}),
\begin{align}
\label{b}
b=\cos\theta_r\sqrt{-\dfrac{g_{\varphi\varphi}}{g_{tt}}}.
\end{align}
At every point on a single orbit, $b$ remains constant. Consequently, we can write it correctly for a particle ejected tangentially from the surface ($\theta_R = \pi/2$) \cite{salmonson1999general,prasanna2002energy},
\begin{align}
\label{cos_theta}
\cos\theta_r=\sqrt{\dfrac{g_{\varphi\varphi(R)}g_{tt}(r)}{g_{\varphi\varphi(r)}g_{tt}(R)}}.
\end{align}
$R$ represents a gravitational source's radius. The rate of energy deposition per unit of time and volume may be expressed as \cite{salmonson1999general}
\begin{align}
\label{dot_q}
\dot{q}=\dfrac{7DG_F^2\pi^3\zeta(5)}{2}\left[KT_{\nu\bar{\nu}}(r)\right]^9\Theta(r),
\end{align}
where $G_F=5.29\times10^{-44}\,\mathrm{cm^2MeV^{-2}}$ is the Fermi constant, 
\begin{align}
 D=1\pm4\sin^2\theta_W+8\sin^4\theta_W,
\end{align}
with the Weinberg angle $\sin^2\theta_W=0.23$. The muon and tau types are represented by the negative sign, whereas electron neutrinos and antineutrinos are represented by the positive sign. The local observer's measurement of the temperature is $T_{\nu\bar{\nu}}(r)$, the neutrino temperature in the neutrino-sphere reads
\begin{align}
\label{Trnu}
T_{\nu\bar{\nu}}(r)\sqrt{g_{tt}(r)}=T_{\nu\bar{\nu}}(R)\sqrt{g_{tt}(R)}.
\end{align}
$\Theta(r)$ is the angular integration factor \cite{lambiase2020effects},
\begin{align}
\label{Theta}
\Theta(r)=\dfrac{2\pi^3}{3}(1-x)^4\left(x^2+4x+5\right),
\end{align}
where
\begin{align}
\label{x}
x=\sin\theta_r=\sqrt{1-\dfrac{g_{\varphi\varphi}(R)g_{tt}(r)}{g_{\varphi\varphi}(r)g_{tt}(R)}},
\end{align}
which can be obtained from Eq.(\ref{cos_theta}). The redshift-related luminosity can be determined as \cite{lambiase2021neutrino}
\begin{align}
\label{Linf}
L_{\nu\bar{\nu},\infty}=g_{tt}(R)L_{\nu\bar{\nu}}(R),
\end{align}
where the neutrino-sphere's luminosity for a single neutrino species is
\begin{align}
L_{\nu\bar{\nu}}(R)=\frac{7}{4}a\pi R^2T_{\nu\bar{\nu}}^4(R),
\end{align}
with the radiation constant $a$. For a single neutrino flavor, the entire amount of local energy deposited in the time unit by the neutrino pair annihilation process can be achieved by integrating Eq.(\ref{dot_q}) from $R$ to infinity,
\begin{align}
\label{dot_Q}
\dot{Q}=4\pi\int_R^{\infty}\dot{q}r^2\sqrt{g_{rr}}\mathrm{d}r,
\end{align}
which can be stated in the way shown below
\begin{align}
\label{dot_Q51}
\dot{Q}_{51}=1.09\times10^{-5}DL_{\nu\bar{\nu},51}^{\frac{9}{4}}R_{6}^{-\frac{3}{2}}\dfrac{\dot{Q}}{\dot{Q}_{\text{Newt}}}.
\end{align}
where $\dot{Q}_{51}$ and $L_{\nu\bar{\nu},51}$ are the total energy deposition and luminosity in units of $10^{51}\,\mathrm{erg/s}$, $R_6\equiv R/10\,\mathrm{km}$, and
\begin{align}
\label{rate_dotQ}
\dfrac{\dot{Q}}{\dot{Q}_{\text{Newt}}}&=3g_{tt}^{\frac{9}{4}}(R)\notag\\
&\quad\times\int_{1}^{\infty}(1-x)^4\left(x^2+4x+5\right)\sqrt{\dfrac{g_{rr}(yR)}{g_{tt}^9(yR)}}y^2\mathrm{d}y,
\end{align}
with $y\equiv r/R$. $\dot{Q}$ will be roughly equivalent to $\dot{Q}_{\text{Newt}}$ in the Newtonian limit, $M\to0$, where $\dot{Q}_{\text{Newt}}$ is obtained from Eq.(\ref{dot_Q51}), resulting in $\dot{Q}/\dot{Q}_{\text{Newt}}=1$.

\begin{figure}
\centering
\includegraphics[scale=0.55]{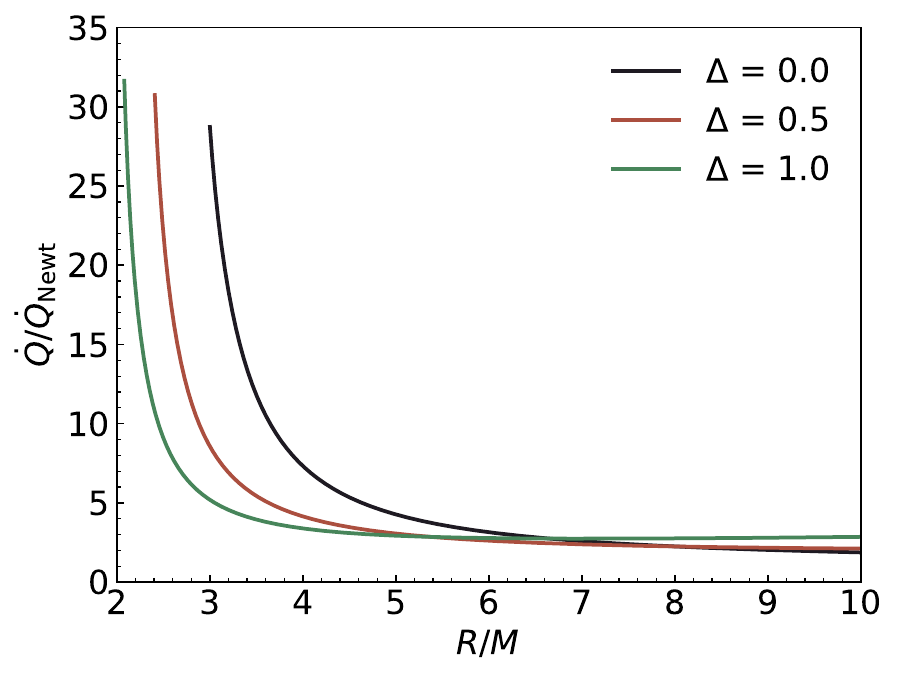}
\caption{The curves of the ratio $\dot{Q}/\dot{Q}_{\text{Newt}}$ as functions of the ratio $R/M$ for parameters $\Delta=0, 0.5$ and $1$.}
\label{fig_FMR}
\end{figure}

Eq.(\ref{rate_dotQ}), which enhances the $\dot{Q}/\dot{Q}_{\text{Newt}}$, is then examined. Here, we adhere to the concepts discussed in the study \cite{salmonson1999general, salmonson2001gamma,salmonson2001neutrino,salmonson2002model}. Salmonson claims that in a near binary system, merging neutron stars may generate gravitational binding energy that may be transformed into internal energy in addition to undergoing relativistic compression and heating in a matter of seconds. Therefore, it's plausible that thermally generated neutrinos might be expelled just before the star falls into a nearby star. When neutrinos ($\nu\bar{\nu}$) emerge from merging stars, they will annihilate to deposit energy, which will then result in the production of electron-positron couples. A strong gravitational field around the star will bend the neutrino paths, boosting the annihilation and scattering rates, as has been noted in the literature. Specifically, when $R/M\sim3-4$ (before to stellar collapse), the $\nu\bar{\nu}$ annihilation deposition energy is increased by a factor of $\dot{Q}/\dot{Q}_{\text{Newt}}\sim\mathcal{O}(8-28)$. We demonstrate in Fig.\ref{fig_FMR} that the existence of a quantum fractal $\Delta$ deforms spacetime, increasing the energy enhancement.

Additionally, we can determine the energy deposition rates corresponding to $R = 3M$ and $4M$ for various fractal parameters $\Delta$ in Tab.\ref{tab_dotQ}, assuming that neutrino pair annihilation is the sole energy source for short GRBs and that the deposited energy is highly efficiently transformed into kinetic energy and photons inside relativistic jets. These findings imply that energy deposition can be enhanced by the effects of quantum fractals and that the neutrino-pair annihilation process in the theory of gravitational corrections beyond GR may effectively contribute to GRB radiation. This may serve as a test site for investigating the non-standard gravity that GR describes.

\begin{figure}
\centering
\includegraphics[scale=0.55]{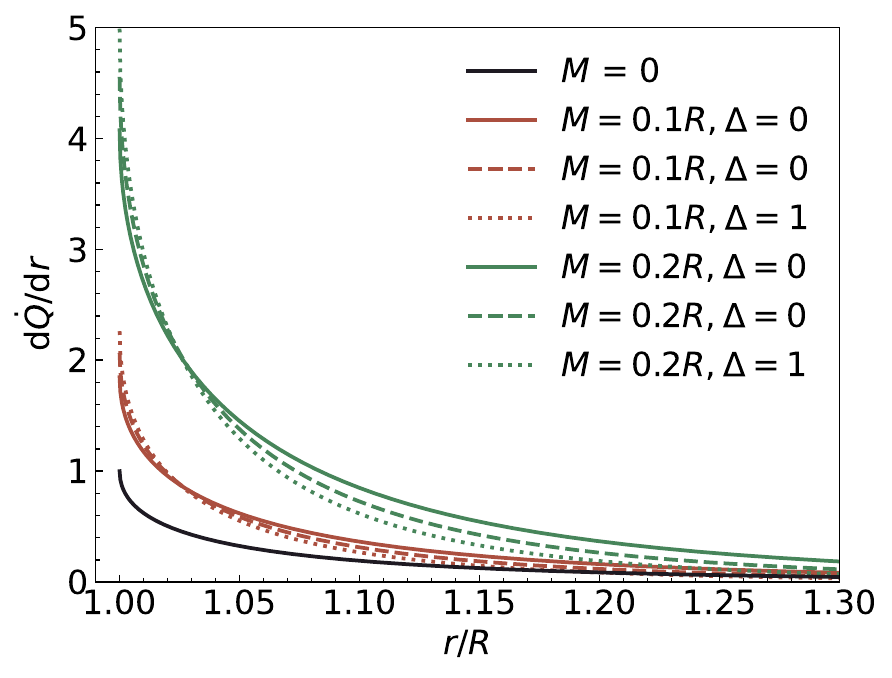}
\caption{The curves of the ratio $\mathrm{d}\dot{Q}/\mathrm{d}r$ as functions of the ratio $r/R$ for parameters $\Delta=0, 0.5$ and $1$.}
\label{fig_dQdr}
\end{figure}

We have also displayed $\mathrm{d}\dot{Q}/\mathrm{d}r$ against radius in Fig.\ref{fig_dQdr}. Additionally, when $M = 0$, we illustrate the Newtonian example for $r/R = 1$ and $\mathrm{d}\dot{Q}/\mathrm{d}r = 1$. It is evident that taking into account GR accelerates the pace of heating. As the $R/M$ values rise, this improvement becomes more noticeable. It's interesting to see that adding the quantum fractal $\Delta$ does not improve every $\mathrm{d}\dot{Q}/\mathrm{d}r$ on $r/R$. The improvement of the heating rate is quite substantial around $r/R\sim1$, as seen in Fig.\ref{fig_dQdr}, but when $r/R$ rises, the parametric $\Delta$ leads it to fall below the GR case or decline more rapidly.

\begin{table}[h!]
\centering
\caption{\label{tabbb}The rate $\dot{Q}$ for different values of $\Delta$ and $R/M$.}
\label{tab_dotQ}
\begin{tabular}{lcc}
\toprule[1pt]
\quad & $R/M$ & $\dot{Q}$ $(\mathrm{erg/s})$ \\
\hline
Newtonian & 0 & $1.50 \times 10^{50}$ \\
\hline
\multirow{2}{*}{$\Delta=0$} & 3 & $4.32 \times 10^{51}$ \\
                            & 4 & $1.10 \times 10^{51}$ \\
\hline
\multirow{2}{*}{$\Delta=0.5$} & 3 & $1.29 \times 10^{51}$ \\
                              & 4 & $0.62 \times 10^{51}$ \\
\hline
\multirow{2}{*}{$\Delta=1$} & 3 & $0.78 \times 10^{51}$ \\
                            & 4 & $0.51 \times 10^{51}$ \\
\bottomrule[1pt]
\end{tabular}
\end{table}

\section{Conclusion}

A quantifiable correlation between $\Delta$ and the accretion disk's energy radiation and energy deposition rate is established for the first time in this research. This work systematically elucidates the ways in which Barrow's fractal black hole geometry \cite{bambi2012code} induces quantum-gravitational spacetime changes that govern high-energy astrophysical events. We show that quantum geometric fluctuations at the black hole horizon notably impact the energy deposition from neutrino-antineutrino pair annihilation and modify the radiation properties of the accretion disk. This is achieved by constructing a modified metric that incorporates the fractal dimension parameter $\Delta$. For $\Delta=1$, it is evident that the innermost stable circular orbit radius moves inward. This geometric reorganization increases the peak radiation flux by $22.5\%$, increases the effective temperature by $62.5\%$, and increases the differential luminosity by around $50\%$ via two processes: orbital velocity distribution renormalization and quantum-enhanced energy dissipation efficiency. The area-law extension of Barrow entropy \cite{barrow2020area}, which views the event horizon as a Koch-type fractal \cite{von2019continuous} with infinite hierarchical layering, is the source of the modification of accretion flow thermodynamics. This approach changes how energy moves through channels by including tiny quantum fluctuations at the small scale, all while keeping the overall volume unchanged.

Quantum-gravitational enhancing effects on neutrino pair annihilation energy deposition rates are induced by the fractal spacetime geometry. Energy deposition rates in key black hole setups, specifically with a ratio of $R/M$ around $3$ to $4$, increase considerably, by factors ranging from $8$ to $28$, when the parameter $\Delta$ is set to $1$. Strong field quantum effects may overcome the requirement that $\Delta\lesssim10^{-3}$ be met for black holes of solar mass to be stable \cite{xia2024upper}. This improvement results from nonlinear coupling in spacetime metric tensors mediating cooperative trajectory bending of neutrinos. Thermodynamic circumstances supporting exponential rise in energy conversion efficiency are created by the redesigned collision angular distribution functions and the resultant local temperature gradients. Demonstrating that spacetime fluctuations resulting from fractal patterns markedly enhance the likelihood of particle interactions under intense gravitational conditions lends strong support to the proposed connection between quantum geometry and large-scale thermodynamic behavior, as described by Barrow entropy.

Mechanistically, the fractal dimension $\Delta$ breaking thermodynamic extensivity is necessary for the physical manifestation of Barrow entropy \cite{petridis2023barrow,capozziello2025barrow, wang2022barrow,abreu2020barrow}. In addition to quantizing the effective potential well of accretion disks, the endlessly nested topology inside the finite horizon volume also creates power-law scaling relations for ISCO radii via geometric-thermodynamic coupling. The distinctive hierarchy of quantum-gravitational corrections in high-energy astrophysical processes is revealed by this multi-scale regulatory mechanism, which places new limitations on the development of neutrino trajectory parameters.

In order to understand jet kinematics in active galactic nuclei, future research should focus on applying the fractal spacetime framework to spinning black hole models. $\Delta$-parameter inversion techniques might be developed using polarimetric measurements of black hole shadows from Event Horizon Telescope data \cite{EHT1,EHT2,EHT3}. Theoretically, determining microscopic interpretations of fractal dimensions within AdS/CFT correspondence \cite{adams2014holographic,pastawski2017code} will depend on investigating the compatibility between Barrow entropy and holographic principles. Gravitational wave spectra from black hole mergers may record ISCO dynamics altered by fractal spacetime, whereas kinematic fingerprints of gamma-ray burst jets and spectral hardness ratios in X-ray binaries may provide observational restrictions on $\Delta$. The geographical distribution of $\Delta$ can be limited by polarization measurements from the Event Horizon Telescope \cite{EHT1,EHT2,EHT3}, and ISCO inward shift characteristics may be present in the LIGO/Virgo merger gravitational wave spectra. These interdisciplinary methods will provide experimental frameworks for using multi-messenger astrophysics to probe quantum spacetime structures.

\vspace{1cm}
\noindent \textbf{Acknowledge}

This work is partly supported by the Shanghai Key Laboratory of Astrophysics 18DZ2271600.

\bibliographystyle{unsrt}
\bibliography{main}

\end{document}